\documentclass[11pt]{article}
\usepackage{axodraw}
\usepackage{epsfig}
\usepackage{amsfonts}
\usepackage{amsmath}
\usepackage{bbm}
\allowdisplaybreaks[1]

\input pix.sty

\renewcommand{\eq}{eq.~}
\renewcommand{\eqs}{eqs.~}
\renewcommand{\se}{sec.~}

\renewcommand{\fig}{fig.~}

\newcommand{\rmO}{{\mathcal{O}}}
\newcommand{\bmu}{\bar\mu}

\def\lsi{\raise0.3ex\hbox{$<$\kern-0.75em\raise-1.1ex\hbox{$\sim$}}}
\def\gsi{\raise0.3ex\hbox{$>$\kern-0.75em\raise-1.1ex\hbox{$\sim$}}}
\newcommand{\lsim}{\mathop{\lsi}}
\newcommand{\gsim}{\mathop{\gsi}}

\newcommand{\nB}{n_\rmii{B}}

 \renewcommand{\nB}[1]{f_\rmii{B{#1}}}
\newcommand{\rmii}[1]{{\mbox{\tiny\rm{#1}}}}

\newcommand{\Tint}[1]{{\hbox{$\sum$}\!\!\!\!\!\!\!\int\,}_{\!\!\!\!\raise-0.9ex\hbox{$\scriptstyle{#1}$}}}
\newcommand{\Tinti}[1]{{{\Sigma}\!\!\!\!\raise0.3ex\hbox{$\int$}_\rmii{${#1}$}}}

\newcommand{\bi}{\begin{itemize}}
\newcommand{\ei}{\end{itemize}}

\newcommand{\hide}[1]{ }

\def\TAsc(#1,#2)(#3,#4,#5)%
{\SetWidth{2.0}\CArc(#1,#2)(#3,#4,#5)\SetWidth{1.0}}
\def\Lwidth{3}

\def\TAgl(#1,#2)(#3,#4,#5){\SetWidth{2.0}\PhotonArc(#1,#2)(#3,#4,#5){\Lwidth}%
{6.283 #3 mul 360 div #4 #5 sub #4 #5 sub mul sqrt mul Tdensity mul}%
\SetWidth{1.0}}
\def\TLgl(#1,#2)(#3,#4){\SetWidth{2.0}\Photon(#1,#2)(#3,#4){\Lwidth}
{#1 #3 sub #1 #3 sub mul #2 #4 sub #2 #4 sub mul add sqrt Tdensity mul}%
\SetWidth{1.0}}
\newcommand{\piC}[1]{\;\parbox[c]{40pt}{\begin{picture}(120,60)(0,-20)
\SetWidth{1.0}\SetScale{0.35} #1 \end{picture}}\;}
\renewcommand{\picc}[1]{\;\parbox[c]{60pt}{\begin{picture}(60,30)(0,0)
\SetWidth{1.0}\SetScale{1.3} #1 \end{picture}}\; }
\def\Lwidth{1.3}

\newcommand{\picg}[1]{\;\parbox[c]{62pt}{\begin{picture}(160,80)(0,-10)
\SetWidth{1.0}\SetScale{0.7} #1 \end{picture}}\;}

\def\ConnectedA(#1,#2,#3){\piC{#1(60,-15)(75,34,146) #2(60,75)(75,214,326)%
 #3(60,60)(20,190,350)%
 \GBoxc(0,30)(10,10){1} \GBoxc(120,30)(10,10){1}%
  }}
\def\ConnectedB(#1,#2,#3){\piC{#1(60,-15)(75,34,146) #2(60,75)(75,214,326)%
 #3(60,60)(60,0)%
 \GBoxc(0,30)(10,10){1} \GBoxc(120,30)(10,10){1}%
  }}
\def\ConnectedC(#1,#2){\piC{#1(60,-15)(75,34,146) #2(60,75)(75,214,326)%
 \GBoxc(0,30)(10,10){1} \GBoxc(120,30)(10,10){1}%
  }}
\def\ConnectedD(#1,#2){\piC{#1(60,-15)(75,34,146) #2(60,75)(75,214,326)%
 \GBoxc(0,30)(10,10){1} \GBoxc(120,30)(10,10){1}%
 \SetWidth{2.0} 
 \Line(55,55)(65,65)%
 \Line(55,65)(65,55)
  }}

 \def\GraphA{\picg{
  \SetWidth{1.5}
  \Photon(0,10)(50,20){3}{6}
  \Photon(50,20)(100,10){3}{6}
  \Photon(50,20)(50,60){3}{5}
  \Photon(0,75)(50,65){3}{6}
  \Photon(51,60)(101,70){2}{6}
  \Photon(50,65)(100,75){2}{6}
  \GCirc(50,42){5}{0}
  \LongArrowArc(50,120)(45,250,290)
  \LongArrowArcn(50,-35)(45,110,70)
  \Text(35,62)[c]{$k \sim 3 T$}
  \Text(35,-2)[c]{$p \sim 3 T$}
 }}
 \def\GraphB{\picg{
  \SetWidth{1.5}
  \Lqu(0,10)(50,20)
  \Lqu(50,20)(100,10)
  \Photon(50,20)(50,60){3}{5}
  \Photon(0,75)(50,65){3}{6}
  \Photon(51,60)(101,70){2}{6}
  \Photon(50,65)(100,75){2}{6}
  \GCirc(50,42){5}{0}
 }}
 \def\GraphC{\picg{
  \SetWidth{1.5}
  \SetWidth{1.5}
  \Laqu(0,10)(50,20)
  \Laqu(50,20)(100,10)
  \Photon(50,20)(50,60){3}{5}
  \Photon(0,75)(50,65){3}{6}
  \Photon(51,60)(101,70){2}{6}
  \Photon(50,65)(100,75){2}{6}
  \GCirc(50,42){5}{0}
 }}
 \def\GraphD{\picg{
  \SetWidth{1.5}
  \SetWidth{1.5}
  \Lsc(0,10)(50,20)
  \Lsc(50,20)(100,10)
  \Photon(50,20)(50,60){3}{5}
  \Photon(0,75)(50,65){3}{6}
  \Photon(51,60)(101,70){2}{6}
  \Photon(50,65)(100,75){2}{6}
  \GCirc(50,42){5}{0}
 }}

\makeatletter \@addtoreset{equation}{section} \makeatother
\renewcommand{\theequation}{\arabic{section}.\arabic{equation}}
\makeatletter
\renewcommand\section{\@startsection {section}{1}{\z@}%
                                   {-5.5ex \@plus -1ex \@minus -.2ex}
                                   {2.3ex \@plus.2ex}%
                                   {\normalfont\large\bfseries}}
\renewcommand\subsection{\@startsection{subsection}{2}{\z@}%
                                     {-3.25ex\@plus -1ex \@minus -.2ex}%
                                     {1.5ex \@plus .2ex}%
                                     {\normalfont\normalsize\bfseries}}
\renewcommand\thesection {\@arabic\c@section}
\renewcommand\thesubsection   {\thesection.\@arabic\c@subsection}
\renewcommand{\@seccntformat}[1]{%
\csname the#1\endcsname.\hspace{1.0em}}
\makeatother

\begin{document}

\flushbottom

\begin{titlepage}

\begin{flushright}
\vspace*{1cm}
\end{flushright}
\begin{centering}
\vfill

{\Large{\bf
 Gravitational wave background from Standard Model physics:
  \\[3mm]
 Qualitative features
}} 

\vspace{0.8cm}

J.~Ghiglieri and 
M.~Laine 

\vspace*{0.8cm}

{\em
 Institute for Theoretical Physics, 
 Albert Einstein Center, University of Bern, \\ 
 Sidlerstrasse 5, CH-3012 Bern, Switzerland\\}
 
\vspace*{0.8cm}

\mbox{\bf Abstract}
 
\end{centering}

\vspace*{0.3cm}
 
\noindent
Because of physical processes ranging from microscopic particle 
collisions to macroscopic hydrodynamic fluctuations, 
any plasma in thermal equilibrium emits 
gravitational waves. For the largest wavelengths the emission rate is
proportional to the shear viscosity of the plasma. In the Standard
Model at $T > 160$~GeV, the shear viscosity is dominated by the most
weakly interacting particles, right-handed leptons, and is relatively
large. We estimate the order of magnitude of the corresponding spectrum 
of gravitational waves. Even though at small frequencies (corresponding 
to the sub-Hz range relevant for planned observatories such as eLISA) 
this background is tiny compared with that from non-equilibrium sources, 
the total energy carried by the high-frequency part of the spectrum is 
non-negligible if the production continues for a long time. We 
suggest that this may constrain (weakly) the highest temperature of 
the radiation epoch. Observing the high-frequency part directly sets 
a very ambitious goal for future generations of GHz-range detectors. 

\vfill

 
\vspace*{1cm}
  
\noindent
June 2015

\vfill

\end{titlepage}

%
\section{Introduction}

Gravitational waves offer a possible way to observe phenomena 
taking place very early in the history of the universe. 
Famously, long-wavelength waves are produced during a period of inflation
(cf.\ e.g.\ refs.~\cite{maggiore,mukhanov}). 
However at higher frequencies gravitational waves can also probe
post-inflationary non-equilibrium phenomena, 
such as preheating~\cite{reheat1}--\cite{reheat2}, 
topological defects~\cite{defect1}--\cite{defect2}, 
bubble dynamics related to a first-order phase 
transition~\cite{bubble1}--\cite{bubble2}, or noisy turbulent 
motion~\cite{noise1}--\cite{noise2}.
Recent numerical simulations start to account for both bubble 
dynamics and the subsequent motion~\cite{simu,simu2}.
For a review concerning post-inflationary sources and
the associated observational prospects, see ref.~\cite{review}.
 
It is well-known
that gravitational waves are being produced in thermal equilibrium 
as well (for an example, see ref.~\cite{sw}). 
In thermal equilibrium particles scatter on each other, 
which implies the presences of forces and accelerations. 
However, for a physical momentum $k > T$, the thermal production 
rate is suppressed by $e^{-k/T}$, 
because the energy carried away by the graviton must be extracted from thermal
fluctuations. Since typical particle momenta are $\sim 3T$ and scatterings
are proportional to the coupling strengths responsible for the interactions, 
it may be assumed that the rate is suppressed by 
$\sim \alpha T^3 e^{-k/T} / m_\rmi{Pl}^2$, 
where $\alpha$ is a fine-structure constant and 
$m^{ }_\rmi{Pl}$ is the Planck mass.
In a weakly-coupled system 
such as the Standard Model, this rate is small.   

On the other hand, a thermal system experiences
also long-wavelength fluctuations not associated with single particle 
states. At the smallest momenta $k \ll T$ these can be called 
hydrodynamic fluctuations~\cite{landau9}. 
We are not aware of an estimate of a corresponding contribution
from equilibrium Standard Model physics to the gravitational wave
background, and one purpose of the present
note is to provide for one. (Very similar physics, 
although boosted by a conjectured turbulent cascade, has recently been 
discussed in ref.~\cite{noise2}.)
In contrast to the non-equilibrium 
phenomena mentioned above, the ``advantage'' of 
a thermal contribution is that it is guaranteed to be present. 
Another purpose of our study is to roughly estimate the production rate
at $k > T$, and to motivate the need for a complete computation. 

In order to be more specific, consider the closely analogous
case of the production rate of photons from a plasma which is neutral
but has electrically charged constituents. 
Even though the  
expectation values of the electromagnetic charge density 
and current vanish, 
$
 \langle {n}^{ }_\rmi{em} \rangle^{ }_{ } = {0}
$,
$
 \langle \vec{J}^{ }_\rmi{em} \rangle^{ }_{ } = \vec{0}
$,
thermal fluctuations do induce charge fluctuations which have a non-zero 
root mean square value: 
\be
 \frac{1}{V} \int^{ }_{\vec{x},\vec{y}\in V}
 \bigl\langle
    n_\rmi{em} (\vec{x}) 
    n_\rmi{em} (\vec{y}) 
 \bigr\rangle^{ }_{ }
 \; = \;
 \int^{ }_{\vec{x} \in V}
 \bigl\langle
    n_\rmi{em} (\vec{x}) 
    n_\rmi{em} (\vec{0}) 
 \bigr\rangle^{ }_{ }
 \; = \; T \chi^{ }_\rmi{em}
 \;, \la{chi_em}
\ee
where $V$ denotes the volume, 
$\chi^{ }_\rmi{em}$ a susceptibility,  
and $\langle ... \rangle$ a thermal expectation value.
The susceptibility
is non-zero even without interactions, for instance for 
a plasma of 
free massless Dirac fermions representing electrons and positrons
it reads $\chi^{ }_\rmi{em} = e^2 T^2/3$, 
$e^2 \equiv 4 \pi \alpha^{ }_\rmi{em}$.
Because of diffusion, the charge fluctuations induce
electromagnetic currents, and currents in turn source photons.
Currents can also directly originate from fluctuations. Assuming 
that the photons produced do not equilibrate as fast as the 
plasma, which is the case for instance for the plasma generated
in heavy ion collision experiments, 
the thermal average
of their production rate can be evaluated. 
A text-book computation shows that the rate 
per unit volume can be expressed as~\cite{text1,text2}
\be
 \frac{{\rm d}\Gamma_\gamma(\vec{k})}{{\rm d}^3\vec{k}}
 \; = \; 
 \frac{1}{(2\pi)^3 2 k}
 \sum_{\lambda} 
 \epsilon_{\mu,\vec{k}}^{(\lambda)}
 \epsilon_{\nu,\vec{k}}^{(\lambda)*} 
 \int_{\mathcal{X}}  
  e^{i\mathcal{K}\cdot\mathcal{X}}
 \big\langle 
    J^{\mu}_\rmi{em} (0) 
    J^{\nu}_\rmi{em} (\mathcal{X}) 
 \big\rangle^{ }_{ }
 \;, \la{photon_rate}
\ee 
where $\mathcal{K} \equiv (k,\vec{k})$, $k \equiv |\vec{k}|$;
$\mathcal{X} \equiv (t,\vec{x})$;  
$\mathcal{K}\cdot\mathcal{X} \equiv k t - \vec{k}\cdot\vec{x}$; 
and $ \epsilon_{\mu,\vec{k}}^{(\lambda)} $ denote polarization vectors. 
For $\vec{k} = k\, \vec{e}_3$, the polarization sum only couples to 
the transverse components $J^{1,2}_\rmi{em}$.

For small $k \ll T$, operator ordering plays no role in \eq\nr{photon_rate}, 
and the fluctuations are also uncorrelated in
space and time~\cite{landau9}. Their 
amplitude is related to diffusion or, equivalently, 
to conductivity ($\sigma$).  
This yields finally 
\be
 \frac{{\rm d}\Gamma_\gamma(\vec{k})}{{\rm d}^3\vec{k}}
 \; \stackrel{k \lsim \alpha^{2}_\rmi{s} T}{\approx} \; 
 \frac{2 T\sigma}{(2\pi)^3 k} 
 \; \sim \; \frac{\alpha^{ }_\rmi{em} T^2 }
 {(2\pi)^3 k\, \alpha^{2}_\rmi{s} \ln (1/\alpha^{ }_\rmi{s})}
 \;, \la{photon_res_1}
\ee
where we inserted the parametric form of the conductivity
of a QCD plasma~\cite{amy1,amy2}.
For large $k \gsim 3 T$, in contrast, the rate originates from particle
scatterings rather than hydrodynamic fluctuations, and has the 
parametric form~\cite{amy3,ak} 
\be
 \frac{{\rm d}\Gamma_\gamma(\vec{k})}{{\rm d}^3\vec{k}}
 \; \stackrel{k \gsim 3 T}{\sim} \; 
 \frac{ \alpha^{ }_\rmi{em}
 \alpha^{ }_\rmi{s} \ln (1/\alpha^{ }_\rmi{s})
 T^2 e^{-k/T} }{(2\pi)^3 k}
 \;.  \la{photon_res_2}
\ee 
In the following we show that results analogous to \eqs\nr{photon_res_1}
and \nr{photon_res_2} apply to gravitational waves, just with the 
replacements $\alpha^{ }_\rmi{em} \to T^2 / m_\rmi{Pl}^2$ and 
$\alpha^{ }_\rmi{s}\to \alpha$. 

Our presentation is organized as follows. 
After deriving an expression for the gravitational wave production
rate in \se\ref{se:prod}, we analyze the structure of the 
energy-momentum tensor correlator for $k \ll T$
in \se\ref{se:tensor}. The quantity parametrizing this structure, 
the shear viscosity, is briefly discussed in \se\ref{se:eta}. 
In \se\ref{se:ll} we turn to the other case $k \gsim 3 T$ and
compute the logarithmically enhanced terms in this regime. 
The results are embedded in a cosmological background
in \se\ref{se:cosmo} and compared with a well-studied non-equilibrium
source in \se\ref{se:order}. Section~\ref{se:concl} offers some 
conclusions and an outlook. 

%
\section{Production rate of gravitational waves from thermal equilibrium}
\la{se:prod}

As a first ingredient, we consider the rate at which energy density
is emitted in gravitational waves. The derivation can be carried out in 
two different ways: by treating gravitons as quantized particles, or
through a purely classical analysis. We start with the first method, 
leading to a result analogous to \eq\nr{photon_rate}. We work first
in Minkowskian spacetime, adding cosmological expansion 
in \se\ref{se:cosmo}.

The linearized equation of motion for the metric perturbation $h^{ }_{ij}$
in the traceless transverse gauge reads
\be
 \ddot{h}^\rmii{TT}_{ij} - \nabla^2 h^\rmii{TT}_{ij} = 
 16 \pi G\, T^\rmii{TT}_{ij}
 \;, \la{eom_GW}
\ee
where $G = 1 / m_\rmi{Pl}^2$.
The right-hand side of this equation plays the role of the electromagnetic
current in the photon case. 
The classical energy associated with gravitational waves 
reads
\be
 E^{ }_\rmii{GW} = \frac{1}{32\pi G}
 \int_{\vec{x}\in V}  
 \, \bigl[ \dot{h}^\rmii{TT}_{ij}(t,\vec{x}) \bigr]^2 
 \;, \la{rho_GW}
\ee
where $V$ is a volume. 
It is well-known that the corresponding energy density cannot be localized. 
However, if we express a free $h^\rmii{TT}_{ij}$ as a usual linear
combination of forward and backward-propagating
plane waves, and omit fast oscillations
$\exp(\pm 2 i \omega t)$, then \eq\nr{rho_GW} can be re-interpreted
as a Hamiltonian with a familiar canonical form: 
\be
 H \equiv
 \langle\!\langle E^{ }_\rmii{GW} \rangle\!\rangle = \frac{1}{64\pi G}
 \int_{\vec{x}\in V}  
 \,  
 \Bigl\{ \bigl[ \dot{h}^\rmii{TT}_{ij}(t,\vec{x}) \bigr]^2 + 
 \bigl| \nabla{h}^\rmii{TT}_{ij}(t,\vec{x}) \bigr|^2
 \Bigr\} 
 \;. \la{rho_GW_2}
\ee
Here $\langle\!\langle ... \rangle\!\rangle$ denotes an average over
an oscillation period. From \eq\nr{rho_GW_2} 
canonically normalized fields can be 
identified as $\hat h^\rmii{TT}_{ij} \equiv h^\rmii{TT}_{ij}/\sqrt{32\pi G}$. 
According to \eq\nr{eom_GW} they are sourced as 
$
 \partial_t^2\hat{h}^\rmii{TT}_{ij} - \nabla^2 \hat{h}^\rmii{TT}_{ij} = 
 \sqrt{8 \pi G}\, T^\rmii{TT}_{ij}
$.
We can now directly overtake \eq\nr{photon_rate} for the production
rate of gravitons, by replacing the polarization vectors 
accordingly. Subsequently, weighting the production rate by the energy
carried by individual quanta, we obtain
\be
 \frac{{\rm d}\rho^{ }_\rmii{GW}}{{\rm d}t\, {\rm d}^3\vec{k}} 
 \; = \; 
 \frac{4\pi G}{(2\pi)^3} 
 \sum_{\lambda} 
 \epsilon_{ij,\vec{k}}^{\rmii{TT}(\lambda)}
 \epsilon_{mn,\vec{k}}^{\rmii{TT}(\lambda)*} 
 \int_{\mathcal{X}}
  e^{i\mathcal{K}\cdot\mathcal{X}}
 \big\langle \,
    T_{ }^{ij} (0) \,
    T_{ }^{mn} (\mathcal{X}) \, 
 \big\rangle^{ }_{ }
 \;. \la{graviton_rate}
\ee 
The sum over the polarization vectors yields
\be
 \sum_{\lambda} 
 \epsilon_{ij,\vec{k}}^{\rmii{TT}(\lambda)}
 \epsilon_{mn,\vec{k}}^{\rmii{TT}(\lambda)*} 
 = 
 \Lambda^{ }_{ij,mn} 
 \equiv 
 \fr12\Bigl(P^{ }_{im} P^{ }_{jn} + P^{ }_{in}P^{ }_{jm}
  - P^{ }_{ij}P^{ }_{mn} \Bigr) 
 \;, \la{proj}
\ee
where $P^{ }_{ij} \equiv \delta^{ }_{ij} - k_i k_j / \vec{k}^2$. 
Choosing 
$
 \vec{k} = k\, \vec{e}_3
$ 
and rotating subsequently the diagonal correlator
$
 \langle \fr12 (T^{11}_{ } - T^{22}_{ }) \,
 (T^{11}_{ } - T^{22}_{ }) \rangle 
$
into the non-diagonal one, we obtain
\be
 \frac{{\rm d}\rho^{ }_\rmii{GW}}{{\rm d}t\, {\rm d}\ln k} 
 \; = \; 
 \frac{8 k^3}{\pi m_\rmi{Pl}^2 } 
 \int_{\mathcal{X}}
 e^{ik(t-z)}
 \big\langle \,
    T^{ }_{12} (0) \,
    T^{ }_{12} (\mathcal{X}) \, 
 \big\rangle^{ }_{ }
 \;. \la{graviton_rate_2}
\ee 

In order to be convinced that \eq\nr{graviton_rate} is correct,
let us repeat the analysis on a purely classical level. Fourier transforming
\eq\nr{eom_GW} with respect to spatial coordinates and denoting the retarded
Green's function related to the time evolution by $\Delta(t,k)$, its
time derivative reads 
$
 \dot{\Delta}(t-t',k) = \theta(t-t') \cos(k(t-t'))
 \;. 
$
Dividing by volume, 
the energy density corresponding to \eq\nr{rho_GW} can then be expressed as 
\be
 \rho^{ }_\rmii{GW} = 
 \frac{8\pi G}{V} \int_{\vec{k}}
 \int_{-\infty}^t \!\!\! {\rm d}t'
 \int_{-\infty}^t \!\!\! {\rm d}t''
 \cos(k(t-t'))\cos(k(t-t''))
 \Bigl\langle
   T^\rmii{TT}_{ij}(t',\vec{k})
   T^\rmii{TT}_{ij}(t'',-\vec{k})
 \Bigr\rangle^{ }_{ }
 \;, 
\ee
where $\int_{\vec{k}} \equiv \int \! {\rm d}^3\vec{k}/(2\pi)^3$.
Following now a standard argument, let us assume that the 
sources switch off before the observation time
$t$. Then the upper bounds of the time integrals can be treated
as independent of $t$, and we can average over fast oscillations
within the integrand:
\ba
 I(t) & \equiv & 
 \Big\langle\!\!\Big\langle
  \int_{-\infty}^t \! {\rm d}t' 
  \int_{-\infty}^t \! {\rm d}t'' \, 
  \cos(k(t-t'))\cos(k(t-t'')) \, \phi(t',t'')
 \Big\rangle\!\!\Big\rangle
 \nn 
  & \simeq &  \fr12
  \int_{-\infty}^t \! {\rm d}t' 
  \int_{-\infty}^t \! {\rm d}t'' \,
 \big\langle\hspace*{-0.8mm}\big\langle
  \cos(k(t'-t'')) + \cos(k(2t-t'-t''))
 \big\rangle\hspace*{-0.8mm}\big\rangle  \, \phi(t',t'')
 \nn 
  & = &  \fr12 
  \int_{-\infty}^t \! {\rm d}t' 
  \int_{-\infty}^t \! {\rm d}t'' \,
 \cos(k(t'-t''))
 \, \phi(t',t'') 
 \;. 
\ea
Taking 
a time derivative and assuming that $\phi$ is a function of the 
time difference\footnote{%
 This can be justified, for instance, if the typical time differences 
 are short (say, reflecting physics much within the horizon) compared 
 with the observation time scale (say, the Hubble time). 
 } 
yields
\ba
 \dot{I}(t) & = &  \fr12
  \int_{-\infty}^t \! {\rm d}t' 
  \, 
  \cos(k(t-t')) \, \bigl[ \phi(t'-t) + \phi(t-t') \bigr]
 \nn
 & = &  
 \fr12 \int_{-\infty}^{\infty} \! {\rm d}\tau \, \cos(k \tau) \, \phi(\tau)
 \; = \; 
 \fr12 \int_{-\infty}^{\infty} \! {\rm d}\tau \, e^{i k \tau}
 \, \frac{ \phi(\tau) + \phi(-\tau) }{2}
 \;.
\ea
Going finally back to configuration
space and making use of translational invariance
in spatial and temporal directions we get 
\be
 \frac{{\rm d}\rho^{ }_\rmii{GW}}{{\rm d}t\, {\rm d}^3\vec{k}} = 
 \frac{4 \pi G}{(2\pi)^3}  
 \int_\mathcal{X}
 e^{i (kt - \vec{k}\cdot\vec{x})} \,
 \Bigl\langle
 \fr12 \bigl\{ 
   T^\rmii{TT}_{ij}(t,\vec{x})
   \, , \, T^\rmii{TT}_{ij}(0,\vec{0})
   \bigr\} 
 \Bigr\rangle^{ }_{ }
 \;. \la{gravi_cl}
\ee
Given that \eq\nr{proj} defines a projection operator to 
the TT modes and that in the 
classical limit operator ordering plays no role, \eq\nr{gravi_cl}
indeed agrees with \eq\nr{graviton_rate} for $k \ll T$. 

%
\section{Correlation function in the tensor channel}
\la{se:tensor}

Having obtained \eq\nr{graviton_rate_2}, the next task is to determine
the shape of the energy-momentum tensor correlator
in momentum space. Here we do this for small 
light-like four-momenta ($\omega, k \lsim \alpha^2 T$), 
returning to the regime $k \gsim 3T$ in \se\ref{se:ll}. 

Consider hydrodynamic fluctuations associated with a local flow
velocity $v^i$ around an 
equilibrium state at a temperature $T$. 
To first order in gradients and in $v^i$,\footnote{%
 Second order terms such as $(e+p) v^i v^j$ are omitted. 
 } 
the energy-momentum tensor has the form
\ba
 T^{0i}_{ } & = & (e+p) \, v^i \;, \la{T0i} \\ 
 T^{ij}_{ } & = & 
 \bigl( p - \zeta \nabla\cdot \vec{v} \bigr) \, \delta_{ }^{ij}
 - \eta \bigl( \partial^{ }_i v^j_{ } + \partial^{ }_j v^i_{ }
 - \fr23 \delta^{ij}_{ } \nabla\cdot \vec{v} \bigr)
 \;, \la{Tij}
\ea
where $e,p,\zeta,\eta$ are the energy density, pressure, 
bulk viscosity, and shear viscosity, respectively. The equation for
energy-momentum conservation asserts that
$
 \partial^{ }_0 T^{0j}_{ } + \partial^{ }_i T^{ij}_{ } = 0
 \; \forall j \in \{ 1,2,3 \}
$.
Let us consider a plane wave perturbation with a momentum vector
$\vec{k} = k\, \vec{e}_3$. Then the equations of motion for the 
transverse velocity components ($\vec{v}_\perp\cdot\vec{k} =0 $) decouple 
from the equations relating $v^3$ and $\partial^{ }_3 p$. The resulting
system is immediately integrated to obtain 
\be
 \vec{v}_{\perp}^{ }(t,\vec{k})
 = \vec{v}_{\perp}^{ }(0,\vec{k}) \, e^{ - \eta k^2 t /(e+p)}
 \;. \la{v_sol}
\ee
We now consider the 2-point correlator 
\be
 \Big\langle \fr12 
 \bigl\{ T^{0i}_{}(t,\vec{k}) , T^{0j}_{}(0,-\vec{k}) 
 \bigr\} \Big\rangle^{ }_{ }
 \;, 
\ee
where the operator ordering is only relevant in the quantum theory.
This correlator is symmetric in $t\to -t$ and has a classical limit. 
Therefore equations \nr{T0i} and \nr{v_sol} lead to a hydrodynamic prediction
for the transverse components ($i',j' \in \{ 1,2 \}$), 
\ba
 & & \hspace*{-2cm}
 \frac{1}{V} \int_{-\infty}^{\infty} \! {\rm d}t \, 
   e^{i \omega t }
 \Bigl\langle
  \fr12 
 \bigl\{ T^{0i'}_{}(t,\vec{k}) , T^{0j'}_{}(0,-\vec{k}) 
 \bigr\}
 \Big\rangle^{ }_{ }
 \nn 
 &  = &  
  \frac{\frac{2 \eta k^2}{e+p}}{ \omega^2 + \frac{\eta^2 k^4 }{ (e+p)^2}}
 \int_{\vec{x}\in V} \!\!\! e^{-i\vec{k}\cdot\vec{x}}
 \Bigl\langle
   T^{0i'}_{}(0,\vec{x}) \, T^{0j'}_{}(0,\vec{0}) 
 \Big\rangle^{ }_{ }
 \;, \la{corr_1}
\ea
where we returned to configuration space for the equal-time correlator. 

Let us take $k$ to be very small, 
and look for the leading term in this limit. In the equal-time correlator
we can send $\vec{k}\to \vec{0}$. Then it equals the susceptibility related
to the total momentum in the $i'$-direction: 
\be
 \int_{\vec{x}\in V} 
 \Bigl\langle
   T^{0i'}_{}(0,\vec{x}) \, T^{0j'}_{}(0,\vec{0}) 
 \Big\rangle^{ }_{T }
 \; = \; \frac{1}{V} 
 \int_{\vec{x},\vec{y}\in V} 
 \Bigl\langle
   T^{0i'}_{}(0,\vec{x}) \, T^{0j'}_{}(0,\vec{y}) 
 \Big\rangle^{ }_{ }
 \; \equiv  \;
 \delta^{i'j'}_{ } \, 
 \chi^{ }_{\vec{p}}
 \;. 
\ee
Even though the average momentum is zero, its susceptibility is non-zero, 
in analogy with \eq\nr{chi_em}:
\be
 \chi^{ }_{\vec{p}} = T (e+p)
 \;. \la{chi_p}
\ee 
Despite including an integral over operator 
correlations at short distances, this exact equation is  
ultraviolet finite just like \eq\nr{chi_em}
(for a rigorous discussion see ref.~\cite{gm1}).

Now, a Ward identity related to energy-momentum conservation 
asserts that
\be
 \omega^2 \, 
 \Bigl\langle
  \fr12 
 \bigl\{ T^{0i'}_{}(\omega,\vec{k}) , T^{0j'}_{}(-\omega,-\vec{k}) 
 \bigr\}
 \Big\rangle^{ }_{ }
 \; = \; 
 k^2 \,
 \Bigl\langle
  \fr12 
 \bigl\{ T^{3i'}_{}(\omega,\vec{k}) , T^{3j'}_{}(-\omega,-\vec{k}) 
 \bigr\}
 \Big\rangle^{ }_{ }
 \;. 
\ee
Therefore, \eqs\nr{corr_1}--\nr{chi_p} can be re-expressed as
\be
 \int_{\mathcal{X}} 
 e^{i (\omega t - k z) }
 \Bigl\langle
  \fr12 
 \bigl\{ T^{3i'}_{ }(\mathcal{X}) , T^{3j'}_{}(0) 
 \bigr\}
 \Big\rangle^{ }_{ }
 \; \stackrel{\omega,k \lsim \alpha^2 T}{ = }  \;
 \frac{2\eta T \omega^2 \delta^{i'j'}}
  {\omega^2 + \frac{\eta^2 k^4 }{ (e+p)^2} }
 \;. \la{kubo_1} 
\ee
Taking $\lim_{\omega\to 0}\lim_{k\to 0}$ from here yields
the well-known Kubo formula for $\eta$.

Of interest to us is {\em not} the correlator of \eq\nr{kubo_1}
(known as the ``shear channel'') but the corresponding correlator
for the spatial components transverse to $\vec{k}$
(known as the ``tensor channel''). It can be argued, 
however, that its functional form is closely related to 
that in \eq\nr{kubo_1}. The tensor components also experience
hydrodynamical fluctuations; but, in our coordinate system 
with $\vec{k} = k\, \vec{e}_3$, these are related to the 
velocity gradients $\partial_1 v^2$ or $\partial_2 v^1$, 
cf.\ \eq\nr{Tij}. 
 These components are decoupled from the equations of motion following 
 from energy-momentum conservation. Therefore, they are not represented
 by smooth differentiable functions responsible for the transfer of 
 hydrodynamic information from one point and time to another; rather, 
 nearby points are uncorrelated, as is the case for generic thermal 
 fluctuations~\cite{landau9}: 
\be
 \Big\langle \fr12 
 \bigl\{ T_{i'j'}^\rmii{TT}(t_1,\vec{x}_1) , T_{k'l'}^\rmii{TT}(t_2,\vec{x}_2) 
 \bigr\} \Big\rangle^{ }_{ }
 \; = \; 
 \Phi^{ }_{i'j'k'l'} \, \delta(t_1-t_2) \, \delta^{(3)}(\vec{x}_1 - \vec{x}_2) 
 \;.
\ee
Consequently a Fourier transform like in \eq\nr{kubo_1} is 
independent of $\omega,k$. Putting finally $\omega = k$ and 
sending $k \to 0$ so that the distinction between spatial
directions disappears, we can fix the coefficient
$\Phi$ through a comparison with \eq\nr{kubo_1}:
\be
 \lim_{k \to 0}
 \int_{\mathcal{X}} 
  e^{ik(t-z)}
 \Bigl\langle
  \fr12 
 \bigl\{ T^{ }_{12}(\mathcal{X}) , T^{ }_{12}(0) 
 \bigr\}
 \Big\rangle^{ }_{ }
 = 2\, \eta\, T 
 \;. \la{kubo2}
\ee
This is the main result that is needed below. We remark that, apart from 
the physical arguments discussed above, the same expression can 
be derived more formally by a linear response analysis related to
a metric perturbation (cf.\ e.g.\ ref.~\cite{ht}).

%
\section{Estimate of shear viscosity at $T > 160$ GeV}
\la{se:eta}

Shear viscosity ($\eta$) is a macroscopic property of a plasma
that originates from 
the microscopic collisions that its constituents are undergoing. It is 
inversely proportional to a scattering cross section and therefore large
for a plasma in which there are some weakly interacting particles. In the 
Standard Model above the electroweak crossover, right-handed leptons
are the most weakly interacting degrees of freedom, changing their
momenta only through reactions mediated by hypercharge gauge fields. 

Omitting for the moment all particle species which equilibrate
faster than right-handed leptons, the shear viscosity can be extracted
from refs.~\cite{amy1,amy2}: 
\be
 \eta \simeq \frac{16 T^3}{g_1^4 \ln(5 T / m^{ }_\rmii{D1})}
 \;,  \la{eta_nll}
\ee
where 
$
 m_\rmii{D1} = \sqrt{11/6}\, g_1 T
$
is the Debye mass related to the hypercharge gauge field. Inserting
$g_1 \sim 0.36$ for the gauge coupling we obtain
\be
 \eta \simeq 400\, T^3
 \;. \la{rough}
\ee
We use this value for order-of-magnitude estimates below. 

If we increase the 
temperature above 160~GeV, the hypercharge coupling $g_1$
grows and the weak and strong couplings $g_2,g_3$ decrease. 
Presumably, the top Yukawa coupling $h_t$ and the Higgs self-coupling
$\lambda$ are also of a similar magnitude. In this situation the analysis
of refs.~\cite{amy1,amy2} should be generalized to include a scalar
field and a more complicated set of reactions. Even though conceptually
straightforward, implementing and solving numerically the corresponding
set of rate equations is a formidable
task and beyond the scope of the present investigation. We note, however,
that the shear viscosity is likely to decrease with increasing $g_1$, 
so that \eq\nr{rough} should represent the most ``optimistic'' 
estimate from the point of view of detecting a thermally emitted
low-frequency gravitational wave background. 

%
\section{Leading-logarithmic production rate at large momentum}
\la{se:ll}

%
\begin{figure}[t]
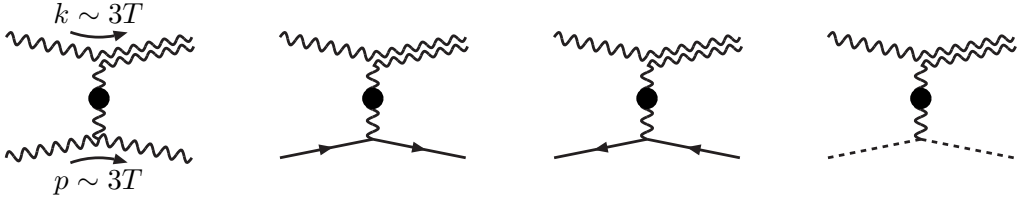


\vspace*{-1.0cm}%

\begin{eqnarray*}
&& 
 \hspace*{-0.7cm}
 \GraphA
 \hspace*{1.25cm}
 \GraphB
 \hspace*{1.25cm}
 \GraphC
 \hspace*{1.25cm}
 \GraphD
 \\[1mm] 
\end{eqnarray*}

\vspace*{-0.8cm}%

\caption[a]{\small 
 Processes leading to a logarithmically enhanced graviton production rate.
 Wiggly lines denote gauge bosons; arrowed lines fermions; dashed lines
 scalars; and a double 
 line a graviton. By $k,p \sim 3T$ we denote typical momenta
 of the scattering particles, whereas the filled blob indicates that the
 vertical rung carries a soft spacelike momentum transfer 
 ($t\sim - \vec{q}_\perp^2 \sim -g^2 T^2$, 
 where $\vec{q}_\perp\cdot\vec{k} = 0$)
 so that the gauge boson needs to be Hard Thermal Loop resummed. 
} 
\la{fig:scat}
\end{figure}
%

Before turning to numerical estimates we wish to complete the qualitative
picture concerning the thermal graviton production rate by considering
the case of ``hard momenta'', $k \sim 3 T$. A full computation of the 
rate in this regime represents a complicated task, similar to the full 
computation of the shear viscosity when all couplings are of the same order
of magnitude, and is postponed to future work.  
In contrast to the shear viscosity, for hard momenta the
result is dominated by the largest couplings, in particular the strong
gauge coupling. If we restrict to logarithmically enhanced terms
(cf.\ \eq\nr{photon_res_2}) then it can be shown that 
only the gauge couplings 
($g_1,g_2,g_3$) contribute at leading order.

An elegant way to determine the logarithmically enhanced terms has been 
discussed in ref.~\cite{bb}, \se{4.2}. Scatterings experienced by soft 
space-like gauge bosons, the vertical rung in \fig\ref{fig:scat}, correspond
to Landau damping, and can be represented within
the Hard Thermal Loop (HTL) effective theory~\cite{htl1,htl2}. Computing
the 2-point correlator of $T^{ }_{12}$ within the HTL theory and noting
that only one of the gauge bosons attaching to the graviton vertex can
be soft at a time, yields (for $q \ll k \sim 3 T$) 
\ba
 \int_{\mathcal{X}}
 e^{ik(t-z)}
 \big\langle \,
    T^{ }_{12} (0) \,
    T^{ }_{12} (\mathcal{X}) \, 
 \big\rangle^{ }_{ }
 & {\approx} & 
 { \nB{}(k) k T }
  \int_{\vec{q}_\perp}^{(\Lambda)} 
  \int_{-\infty}^{\infty} \! \frac{{\rm d} q_\parallel}{2 \pi}
  \biggl\{ 
    \frac{ \rho_\rmii{T}(q_\parallel,\vec{q}) }{q_\parallel} 
   - \frac{ \rho_\rmii{E}(q_\parallel,\vec{q}) }{q_\parallel} 
  \biggr\} 
  \frac{q_\perp^4}{ q_\perp^2 + q_\parallel^2}
 \nn 
 & = & 
 { \nB{}(k) k T }
  \int_{\vec{q}_\perp}^{(\Lambda)} 
  \biggl( \frac{1}{q_\perp^2} - 
  \frac{1}{q_\perp^2 + m_\rmii{D}^2} \biggr)\,  \frac{q_\perp^2}{2} 
 \;, \la{ll}
\ea
where $\nB{}$ is the Bose distribution; 
$q_\parallel \equiv \vec{q}\cdot\vec{k} / k$; 
$m_\rmii{D}^2$ is a Debye mass squared;  
$\Lambda$ indicates that this treatment only applies
to soft modes $q_\perp \ll 3 T$; and $\rho^{ }_\rmii{T/E}$
are spectral functions corresponding to the ``transverse'' and
``electric'' polarizations, respectively.  
In the last step we made use of a sum rule for the 
HTL-resummed gluon propagator that has been derived in refs.~\cite{agz,sch}.

The integral in \eq\nr{ll} happens to be identical to that appearing in 
the context of the jet quenching parameter $\hat{q}$ in QCD~\cite{sch}. 
Carrying it out and inserting the result into \eq\nr{graviton_rate_2}, 
we obtain
\be
 \frac{{\rm d}\rho^{ }_\rmii{GW}}{{\rm d}t\, {\rm d}\ln k} 
 \; = \; 
 \frac{2 k^4 T \nB{}(k)}{\pi^2 m_\rmi{Pl}^2 } 
 \biggl\{  
 \sum_{i=1}^3 d_i\, m_\rmii{D$i$}^2 \ln \frac{5 T}{ m^{ }_\rmii{D$i$}}
   + \rmO\Bigl( g^2 T^2 \chi\Bigl( \frac{k}{T} \Bigr) \Bigr)
 \biggr\}
 \;, \la{ll_2}
\ee
where $d_1 \equiv 1$, $d_2 \equiv 3$, $d_3 \equiv 8$; 
$m^{ }_\rmii{D$i$}$ is the Debye mass corresponding to the gauge 
group U(1), SU(2) or SU(3), respectively; 
$g^2 \in \{ g_1^2, g_2^2, g_3^2, h_t^2 \}$; and the ultraviolet
scale within the 
logarithm has been (arbitrarily) taken over from \eq\nr{eta_nll}. 
The Debye masses read
$
 m^2_\rmii{D1} = 11 g_1^2 T^2/6
$,
$
 m^2_\rmii{D2} = 11 g_2^2 T^2/6
$,
and
$
 m^2_\rmii{D3} = 2 g_3^2 T^2
$.
Because of the largest multiplicity, 
the result is dominated by the QCD contribution.
We note that a similar computation for $t$-channel
fermion or Higgs exchange does not lead to logarithmic enhancement. 

%
\section{Embedding the result in cosmology}
\la{se:cosmo}

\begin{figure}[t]


\centerline{%
 \epsfysize=8.0cm\epsfbox{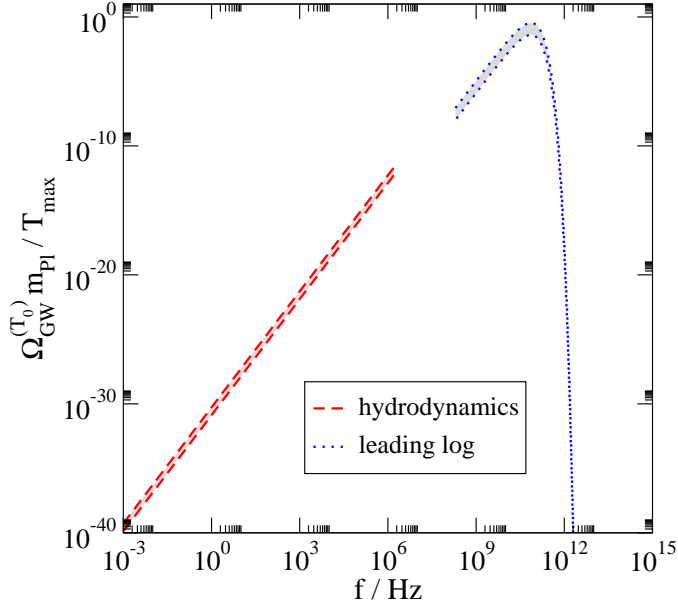}%
}

\caption[a]{\small
The result from \eq\nr{graviton_rate_5}, multiplied by 
$m^{ }_\rmi{Pl} / T_\rmi{max} $, as a function of the 
present-day frequency. 
 The maximum of the power lies in the range $k \sim T_\rmi{max}$
 at $T = T_\rmi{max}$, and at $k \sim T_0$ at $T = T_0$.
The hydrodynamic and leading-log results correspond to the two limits
shown in \eq\nr{phi}, with the band originating
from varying $\hat{\eta} = 100...400$ in the hydrodynamic prediction
and from varying the constant $\rmO(1)$ within the range 0...10
in the leading-logarithmic result. 
The couplings were fixed at a scale $\bmu = \pi T$ with $T \simeq 10^6$~GeV: 
$g_1^2 \approx 0.13$, $g_2^2 \approx 0.40$, $g_3^2 \approx 1.0$.
For obtaining the current day energy fraction the result
needs to be multiplied by $\Omega^{ }_\rmi{rad} \sim 5\times 10^{-5}$.
The eLISA sensitivity peaks at $f\sim 10^{-2}...10^{-3}$~Hz.
}

\la{fig:llog}
\end{figure}

Combining \eqs\nr{graviton_rate_2}, \nr{kubo2} and \nr{ll_2}, we get 
\be
 \frac{{\rm d}\rho^{ }_\rmii{GW}}{{\rm d}t\, {\rm d}\ln k} 
 \; = \; 
 \frac{16 k^3 \eta T}{\pi m_\rmi{Pl}^2 } \, \phi\Bigl(\frac{k}{T}\Bigr)
 \;. \la{graviton_rate_3}
\ee 
This applies in a local Minkowskian frame. The function $\phi$, 
\be
 \phi\Bigl(\frac{k}{T}\Bigr) \; \simeq \; 
 \left\{ 
  \begin{array}{ll}
    \displaystyle 1 & \;,\quad  k \lsim \alpha^2 T \\
    \displaystyle \frac{  k \nB{}(k) }{8\pi\eta}
   \sum_{i=1}^3  d_i\, m_\rmii{D$i$}^2 
   \biggl( \ln \frac{5 T}{ m^{ }_\rmii{D$i$}} + \rmO(1) \biggr)
    & \;, \quad k \gsim 3 T
  \end{array}
  \right. 
  \la{phi}
\ee
is quantitatively correct at $k \lsim \alpha^2 T$ 
whereas  at $k \gsim 3T$ it only
represents the qualitative structure (in particular the coefficient ``5''
inside the logarithm
is but a convention, and there could be substantial non-logarithmic 
contributions from $h_t^2$ or from $\rmO(g)$-suppressed effects 
like in the case of the jet quenching
parameter $\hat{q}$~\cite{sch}). We would now like to 
re-express the result in an expanding cosmological background, and 
subsequently obtain numerical estimates. As our reference
temperature we take that corresponding to the electroweak crossover
in the Standard Model, $T_0 \equiv 160$~GeV
(cf.\ e.g.\ refs.~\cite{at2,mm}).  

The basic equations needed from cosmology are (for a flat spatial geometry)
\be
 H= \frac{\dot{a}}{a} = \sqrt{\frac{8\pi e}{3}} \frac{1}{m^{ }_\rmi{Pl}}
 \;, \quad
 \frac{a(t)}{a(t_0)} = \biggl[ \frac{s(T^{ }_0)}{s(T)} \biggr]^{\fr13}
 \;, \la{entropy} 
\ee
where $H$ is 
the Hubble rate, 
$a$ is the scale factor, 
and $s(T)$ is the entropy density. 
Combining the two equations in \eq\nr{entropy}, 
the relation between time and temperature can be expressed as
\be
 \frac{{\rm d}T}{{\rm d}t} = - T H(T) 3 c_s^2(T)
 \;,
\ee
where $c_s$ is the speed of sound, $c_s^2(T) = p'(T)/e'(T)$. 
The energy density carried by gravitational waves is of the form
$
 \rho^{ }_\rmii{GW}(t) = \int_\vec{k} k \, f(t,k)
$, 
where $f$ is a phase space distribution. Making use of the known 
evolution equation for $f$ in an expanding background, the 
energy density can be seen to evolve as
\be
 (\partial_t + 4 H) \rho^{ }_\rmii{GW}(t) = \int_\vec{k} R(T,k)
 \;, \la{evol1}
\ee
where $R(T,k) = 32\pi \eta T \phi(k/T) /m_\rmi{Pl}^2 $
in the notation of \eq\nr{graviton_rate_3}. Given that
$(\partial_t + 3 H)s = 0$, the factor $4H$ 
can be taken care of by
normalizing $\rho^{ }_\rmii{GW}$ by $s^{4/3}$. Subsequently 
the equation can be integrated, by assuming that at an initial
time $t^{ }_\rmi{min}$ (corresponding to a maximal temperature
$T^{ }_\rmi{max}$) there were no (thermally produced) 
gravitational waves present:  
\be
 \frac{\rho^{ }_\rmii{GW}(t_0)}{s^{4/3}(t_0)}
 = 
 \int_{t^{ }_\rmi{min}}^{t_0}
  \! {\rm d}t \int_\vec{k} \frac{R(T,k)}{s^{4/3}(t)}
 = \int_{T_0}^{T_\rmi{max}} \! {\rm d}T \int_\vec{k}
 \frac{R(T,k)}{T H(T) 3 c_s^2(T) s^{4/3}(T)}
 \;. \la{evol2}
\ee
Taking into account that momenta redshift as 
$
 {k}(t) = {k}_0\, a(t_0) / a(t)
$
and expressing the momentum space integrals in terms of $k_0$ finally yields
\ba
 \Omega^{ }_\rmii{GW}(k_0) & \equiv & \frac{1}{e(T_0)} 
 \frac{{\rm d}\rho^{ }_\rmii{GW}}{{\rm d}\ln k_0} 
 \nn 
 & = & 
 \frac{8 k_0^3 s^{1/3}(T_0)}{m_\rmi{Pl}^{ } \sqrt{6\pi^3} e(T_0)}
 \int_{T_0}^{T_\rmi{max}} \! {\rm d}T \, 
 \frac{\eta(T) }{ c_s^2(T) s^{1/3}(T) e^{1/2}(T)  } 
 \, \phi\Bigl( \frac{k_0}{T}
 \Bigl[ \frac{s(T)}{s(T^{ }_0)} \Bigr]^{\fr13}
  \Bigr)
 \;,
\ea
where we also inserted $H$ from \eq\nr{entropy}.
If we approximate $c_s^2\approx 1/3$; assume all 
thermodynamic functions to scale with their dimension
($s = \hat{s} T^3, \eta = \hat{\eta} T^3, e = \hat{e} T^4$, 
with $\hat{s}, \hat{\eta}, \hat{e}$ roughly constants at $T > T_0$);   
and consider $T^{ }_\rmi{max} \gg T_0$, then 
\be 
 \Omega^{ }_\rmii{GW}(k_0)
 \simeq  
 \frac{24 \hat{\eta}}{\sqrt{6\pi^3 \hat{e}^3}} \,
 \frac{T_\rmi{max} }{m^{ }_\rmi{Pl} } \,
 \frac{ k_0^3 }{ T_0^3  }
 \, \phi\Bigl( \frac{k_0}{T_0}  \Bigr)
 \;. \la{graviton_rate_5}
\ee
This result is plotted in \fig\ref{fig:llog}, 
after a redshift of $k_0/T_0$ to a current-day frequency.

For a given mode $k_0$, production starts at a maximal temperature when 
the argument of $\phi$ is of order unity; 
since the entropy density roughly scales with $T^3$ for $T > T_0$, this poses
no particular constraint if we restrict to $k_0 \lsim T_0$. 
The horizon radius of a given period redshifts, 
and we are interested in causal physics taking place within the horizon. 
For instance, a temperature leading to a horizon radius comparable
to the planned eLISA~\cite{eLISA} arm length, $\sim 10^6$~km, 
is $T_\rmi{max}\sim 10^6$~GeV. Let us take this as an example. 
Inserting $\hat{\eta}\sim 400$, $\hat{e} \sim 35$, 
$T^{ }_\rmi{max} \sim 10^6$~GeV into \eq\nr{graviton_rate_5}, 
we thus get
\be
  \Omega^{ }_\rmii{GW}(k_0) \sim 3 \times 10^{-13}
 \,
 \times \frac{T^{ }_\rmi{max}}{10^6\,\mbox{GeV}} 
 \times 
 \, \frac{ k_0^3}{T_0^3}
 \, \phi\Bigl( \frac{k_0}{T_0}  \Bigr)
  \;. \la{final}
\ee
The Hubble radius ($H^{-1}$) of the electroweak epoch ($T=T_0$) corresponds
to $\sim 10^{10}$~km today, so if we consider wavelengths
extending up to $\sim 10^6$~km, then 
$k_0 \sim \alpha H$ with $\alpha \sim 2\pi\times 10^4$. 
In this situation $k_0/T_0$ can be estimated as
\be
 \frac{k_0}{T_0} \sim \frac{\alpha H(T_0)}{T_0} =
\sqrt{\frac{8\pi\hat{e}}{3}} \frac{T_0}{m^{ }_\rmi{Pl}} \,  \alpha 
 \sim 2 \times 10^{-16}\,\alpha 
 \;. \la{problem}
\ee
Inserting into \eq\nr{final} we find a very small energy fraction. 
However the fraction is larger if we consider the total amount of energy
in gravitational waves, which originates dominantly from $k_0\sim T_0$; 
this energy is constrained to be 
below that corresponding to one equilibrated relativistic degree 
of freedom~\cite{Smith:2006nka,Henrot-Versille:2014jua}, 
$\int_{\ln k_0} \Omega^{ }_\rmii{GW} \ll 1/100$. We return to this
consideration around \eq\nr{total}, but first compare
the infrared part with non-equilibrium processes.

%
\section{Order-of-magnitude comparison with a non-equilibrium source}
\la{se:order}

In order to get a qualified 
impression about the magnitude of the thermal background, 
consider the well-studied case of a  
first-order phase transition at the electroweak epoch. As before
$\Omega^{ }_\rmii{GW}$ denotes the ratio of the energy densities of 
gravitational waves and radiation at $T_0 \sim 160$~GeV. 
Today, the energy density in radiation corresponds to 
$\Omega^{ }_\rmi{rad} \sim 5\times 10^{-5}$, and the projected
eLISA sensitivity is $\Omega^{ }_\rmii{eLISA} \sim 10^{-11}$. 
So, in order to be detectable, we hope to find a signal in 
the range $\Omega^{ }_\rmii{GW} \sim 2\times 10^{-7}$
at the electroweak epoch. 

Apart from the overall magnitude, an important feature of any observatory
is that its sensitivity peaks in a certain frequency range. For eLISA, 
this is $f \sim (10^{-3}\, ...\, 10^{-2})$~Hz.  
This corresponds to a distance scale 
$\ell^{ }_\rmii{B} \sim (10^{-3}\, ...\, 10^{-2})\, \ell^{ }_\rmii{H} $ 
in terms of the horizon radius of the electroweak epoch, 
where we have defined 
$
  \ell^{ }_\rmii{H} \equiv H^{-1} 
$. 

Now, the overall signal from a first-order transition is traditionally
argued to be of the form~\cite{bubble1}--\cite{bubble2} 
\be
 \Omega^{ }_\rmii{GW} \sim v_w^n\, \kappa^2 
 \Bigl( \frac{L}{e} \Bigr)^2 
 \Bigl( \frac{\ell^{ }_\rmii{B}}{\ell^{ }_\rmii{H}} \Bigr)^2
  \;, \la{estimate}
\ee
where $v_w^{ }$ is a bubble wall velocity; 
$n > 0 $; 
$L$ is the latent heat released in the transition; 
$\kappa < 1$ parametrizes the efficiency at which energy is converted into
gravitational waves \cite{Kamionkowski:1993fg}; 
and $\ell^{ }_\rmii{B}$ is the typical bubble separation. 
The spectrum peaks at momenta corresponding to the 
bubble separation scale, $k_\rmi{max} \sim 2\pi / \ell^{ }_\rmii{B} $.
For $k < k_\rmi{max}$, a behaviour $\sim k^3$ has been 
found, whereas at $k > k_\rmi{max}$ the spectrum falls off, 
perhaps as $1/k$ or $1/k^2$~\cite{bubble2}. A recent numerical 
study shows that the complicated dynamics following bubble collisions 
continues for a long time and thereby boosts \eq\nr{estimate} by a factor
$\sim \ell^{ }_\rmii{H} / \ell^{ }_\rmii{B}$, with the price 
of a more rapid decay of the power spectrum at large $k$~\cite{simu}. 

If we optimistically insert 
$\ell^{ }_\rmii{B} \sim 0.01 \ell^{ }_\rmii{H}$ into \eq\nr{estimate}, 
it still remains a challenge to get $\Omega^{ }_\rmii{GW} \sim 10^{-7}$
out. Basically, a very large latent heat $L/e \gg 0.01$ would be needed. 
This requires a drastic modification of the Higgs sector, which
normally carries just a handful of degrees of freedom 
in comparison with $\sim 100$ contributing to $e$. However, 
with somewhat less drastic assumptions, and including a boost 
by $\sim \ell^{ }_\rmii{H} / \ell^{ }_\rmii{B}$ from ref.~\cite{simu}, 
numbers like $\Omega^{ }_\rmii{GW}\sim 10^{-10}$ could be obtained, 
which is also of interest for future generations of observatories. 

In any case, inserting 
$\alpha \sim 2\pi \ell^{ }_\rmii{H}/\ell^{ }_\rmii{B} \sim 10^{3-4}$
and $T^{ }_\rmi{max} \sim 10^6$~GeV
into \eqs\nr{final}, \nr{problem}, the thermal background would
be $\sim$ 40 orders of magnitude below the desired level at the 
peak eLISA frequency (cf.\ also \fig\ref{fig:llog}). 
What is significant about the 
thermal background, though, is that it continues to grow
with $k$ for another more than 10 decades, and therefore eventually
overtakes the decaying non-equilibrium signal
at short distance scales. In fact, originating as it does 
from fluctuations at the scale $k \sim 3 T^{ }_\rmi{max}$ 
and red-shifting as dictated by entropy conservation,  
the peak power is in the range 
$k \sim 3 T^{ }_\rmi{dec} (3.9 / 106.75)^{1/3} \sim T^{ }_\rmi{dec}$
at  the time of photon decoupling, and 
in the microwave range today.
Therefore it falls in the range of recently conceived
high-frequency 
experiments~\cite{Cruise:2006zt}--\cite{Cruise:2012zz}.

%
\section{Conclusions and outlook}
\la{se:concl}

We have estimated the magnitude and shape of the 
gravitational wave background that is produced by Standard Model
physics during the thermal history of the universe until 
the temperature $T_0 \approx 160$~GeV corresponding to the electroweak
crossover, cf.\ \eqs\nr{phi}, \nr{graviton_rate_5} and 
\fig\ref{fig:llog}. 
The infrared part could have been 
of potential interest in that the forthcoming eLISA 
experiment is probing sub-Hz frequencies with unprecedented precision. 
Unfortunately, we have found that in this range the thermally produced
gravitational wave signal is many orders of magnitude below the 
observable level, cf.\ \se\ref{se:order}.

In general, the thermally produced gravitational wave background resembles
a bit the blackbody spectrum of photons and neutrinos. Its shape is not the
same because gravitational waves never equilibrate. Therefore the shape has
to be determined by a dynamical computation which has not been
carried out even at full leading order. 
Nevertheless, it is already clear from a leading-logarithmic estimate
that the peak of the power today lies in the same microwave domain as for 
photons and neutrinos (cf.\ \fig\ref{fig:llog}). 
Therefore, the best observational prospect lies with high-frequency 
(0.1 -- 4.5~GHz)
experiments~\cite{Cruise:2006zt}--\cite{Cruise:2012zz}.
At the current stage it seems challenging to reach a sensitivity 
below $\Omega^{ }_\rmii{GW}\sim 10^{-5}$~\cite{cq3}, 
whereas an optimistic theoretical expectation would be 
\be
 \Omega^{ }_\rmii{GW} \sim \Omega^{ }_\rmi{rad} \times \frac{1}{100} 
 \times \Bigl( \frac{4.5~\mbox{GHz}}{100~\mbox{GHz}} \Bigr)^3
 \sim 10^{-11}
 \;, 
\ee
where $1/100$ corresponds to the maximally allowed fraction
in gravitational waves at the electroweak epoch when all 
degrees of freedom are relativistic, and 100~GHz to the 
frequency associated with generic blackbody radiation. 
So, there is surely a long way to go till detection. 

There is, however, one consideration which can already be carried out. 
Indeed, unlike neutrinos, gravitational waves {\em must not} 
carry as much energy density as one relativistic degree of 
freedom~\cite{Smith:2006nka,Henrot-Versille:2014jua}. This constrains
the total energy density stored in them and, given that the production
rate peaks at the maximal temperature, the maximal temperature reached. 
The total energy density corresponding to \eq\nr{graviton_rate_5} can 
be estimated as  
\be
 \int \! {\rm d}\ln\! k_0 \, \Omega^{ }_\rmii{GW}(k_0) 
 \; \simeq \;
 \frac{24 \hat{\eta}}{\pi \sqrt{6\pi \hat{e}^3}}
 \frac{T_\rmi{max}}{m^{ }_\rmi{Pl} T_0^3}
 \int_0^\infty \! {\rm d}k_0 \, k_0^2 
 \, \phi\Bigl( \frac{k_0}{T_0}  \Bigr)
 \; \simeq \; 
 \frac{24 }{\pi \sqrt{6\pi \hat{e}^3}}
 \biggl( 8 \ldots \frac{\hat{\eta}}{3} \biggr)
 \frac{T_\rmi{max}}{m^{ }_\rmi{Pl}}
  \;, \la{total}
\ee
 where we varied $\phi$ between two limits: the factor 8 originates if 
 we adopt the form of $\phi$ appearing on the second line of \eq\nr{phi}, 
 setting the unknown constant to zero and the couplings to values
 mentioned in the caption of \fig\ref{fig:llog}, 
 whereas the factor $\hat{\eta}/3$ 
 originates if we use the first line of \eq\nr{phi} and cut off the integral 
 at $k_0 = T_0$: $\int_0^{T_0} {\rm d} k_0 \, k_0^2 = T_0^3/3$.
According to Planck data~\cite{planck} only a small fraction of 
a relativistic degree of freedom beyond those in the Standard Model
can be permitted, so at $T_0 \sim 160$~GeV we must require
\be
 \frac{24 }{\pi \sqrt{6\pi \hat{e}^3}}
 \biggl( 8 \ldots \frac{\hat{\eta}}{3} \biggr)
 \frac{T_\rmi{max}}{m^{ }_\rmi{Pl} }
 \ll 
 \frac{1}{100}
 \;.
\ee
Inserting $\hat{e}\sim 35$, $\hat{\eta}\sim 400$
we obtain $T_\rmi{max} \lsim 10^{17} ... 10^{18}$~GeV. This is not  
a very strong constraint,\footnote{%
 In particular, within the standard inflationary paradigm, reheating 
 temperatures above $\sim 10^{16}$~GeV are considered all but excluded. 
 }
but the estimate could be sharpened
with more knowledge about the function $\phi$. 

To summarize, a determination of the function $\phi$, 
defined by \eq\nr{graviton_rate_3}, beyond the leading-logarithmic
terms that we have obtained here, 
seems to pose an interesting problem.  
This computation represents 
a well-defined challenge in thermal field theory, 
analogous to that for the photon production rate from 
a QCD plasma~\cite{amy3,ak} or the right-handed neutrino production
rate from a Standard Model plasma~\cite{bb,gl}. 
It is technically more challenging, because every single
particle species carries energy and momentum, and therefore 
we leave the practical implementation to future works.  
(A somewhat related computation, but for the off-shell
kinematics $k=0$, $\omega\gsim T$, 
has been presented in ref.~\cite{vz}.)

%
\section*{Acknowledgements}

We thank M.~Hindmarsh for helpful discussions. 
This work was partly supported by the Swiss National Science Foundation
(SNF) under grant 200020-155935.

%
\appendix
\renewcommand{\thesection}{Appendix~\Alph{section}}
\renewcommand{\thesubsection}{\Alph{section}.\arabic{subsection}}
\renewcommand{\theequation}{\Alph{section}.\arabic{equation}}

%

%

\end{document}